# Creating optical centrifuge for rapid optical clearing and separation

XIONGGUI TANG, * YI SHEN AND YANHUA XU

*Department of Physics, Key Laboratory of Low Dimensional Quantum Structures and Quantum Control of Ministry of Education, Hunan Normal University, Changsha, 410081, P.R. China*
*\* tangxg@hunnu.edu.cn,*

**Abstract**: Optical centrifuge has emerged as a promising tool for achieving centrifuge motion of particles in many fields. Herein, we report a novel optical centrifuge, as driven by optical lateral force arising from transverse phase gradient of light in holographic optical tweezers. It has remarkable advantages, including rapid speed, small footprint, flexible design and easy control. The optical clearing and sorting as application examples have been experimentally demonstrated by using Au and polystyrene nanoparticles. The results show that the optical centrifuge exhibits excellent performance, while used as optical clearing and optical sorting. For instance, its centrifuge velocity in optical clearing can reach its peak speed of 120 μm/s in a short time, which is highly desirable for optical manipulation of various particles in microfluidic systems. This study opens a way for realizing novel functions of optical manipulation including optical clearing and optical sorting.

**Keywords**: optical centrifuge; optical clearing; optical sorting; phase gradient profile, optical tweezers.

## 1. Introduction

Nowadays, conventional centrifuges, which apply a centrifugal force to induce particles to run away along tangential direction, are widely used in many fields such as biology, chemistry and medicine [1]. In the past two decades, the lab-on-chip (LOC) devices have gained a rapidly increasing attention, which have ability of biological sample processing, manipulation and analysis on small scale [2-9]. Its attractive advantages include miniaturization, integration and automation in laboratory processes from basic operations to various biochemical reactions. For instance, separation and sorting of cells or particles from heterogeneous mixtures is a critical task in sample preparation in LOC systems. However, the commercial centrifuges are generally bulky and requires electricity-powered, which can't be used in microfluidic systems. Consequently, different proposals have been recently proposed [10-18], most of which are based on inertial microfluidic technique or hydrophoresis effect. However, there exists several disadvantages, such as complex structure, large volume, low throughput, and high cost. As a result, developing optical centrifuges for rapidly and conveniently creating centrifuge motion of particles in microfluidic system is very highly preferred. In addition, the target particles are always collided by the unwanted particles in optical trapping system,, which seriously affect manipulation of the trapped particles, so optical centrifuge would be a promising tool as optical clearing for rapidly removing these unwanted particles.

Previously, optical centrifuges have been demonstrated by using vortex beam or circularly polarized beam [19-21], which are mostly used in separation and rotation of molecules in gas environment, but it is not suitable for centrifuge motion of nanoparticles in fluid environment. The reason is that the drag force in fluid environment is much more than one in the gas environment. To date, optical centrifuges for achieving optical manipulation of nanoparticles in fluid environment, have not been reported yet.

In this work, we report a new type of optical centrifuge, by smartly employing the optical lateral force arising from phase gradient profiles. This strategy is very attractive, including flexible design, easy control, and rapid speed. Specially, its velocity can reach 120 μm/s in a short time. More importantly, the proposed optical centrifuge can not only be used for optical clearing, but also be applied in optical sorting in optical trapping systems. This study provides a new means for achieving optical centrifuge motion in aqueous solution and developing new function of optical manipulation.

## 2. Principle and design

In fluid environment, the particles trapped by vortex beam always rotate along circle orbit [22-26], and won't run away from original orbit, even if the driving force arising from phase gradient increases. The reason is that centripetal force is large enough to maintain the circle motion of particle, which originates from viscosity force of fluids, as illustrated by $f_2$ in Fig. 1a. In addition, optical force $f_3$ arising from phase gradient profiles are always counteracted by drag force $f_1$, which leads to approximate uniform speed motion in circle orbit. In order to achieve the centrifuge motion, the centripetal force should be reduced by adding another optical force, as depicted by $f_4$ in Fig.1b, which is along radial direction. The other optical force $f_3$ is used to drive particle to generate angular velocity. Similarly, the drag force $f_1$ is proportional to the velocity. The drag force $f_2$ always varies, because of its curve motion of particle, and its magnitude are variable with velocity and radius of curve at this point.

Obviously, the intensity gradient force isn't suitable for achieving such complex motion in the large operation area. Naturally, the optical force induced by phase gradient profile is a considerable candidate for the centrifuge motion. Apparently, its phase profile at optical manipulation should have appropriate gradient in the both radial direction and tangential direction. Consequently, its phase profile can be obtained by superimposing two phase profiles, as depicted in Fig. 1c. It can be described as below,

$$\varphi(r,\theta) = l\theta + kr \qquad (1)$$

where $r$ and $\theta$ stand for radial and angle position in polar coordinate system; $l$ and $k$ denote azimuthal and radial index. Additionally, it needs to point out that its intensity profile should be nearly uniform in the large operation area. In the following, its hologram needs to be designed to achieve the desired phase and intensity distribution as well as possible. However, it is a challenging task, i.e. simultaneously generating the desirable intensity and phase profile at the large area is a challenging task. In this case, the holographic optical system is employed, and its hologram is designed by using our proposed method [27]. In addition, the spherical wave-front has been added into its hologram to effectively improve its intensity and phase profiles.

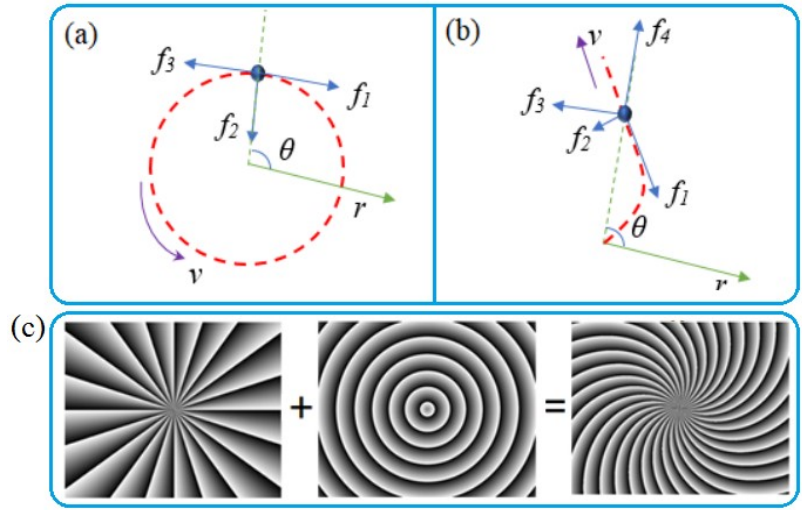

**Fig. 1.** Force analysis of particle motion in fluid environment, (a) circle motion, and (b) centrifuge motion. (c) The desirable phase profile at optical manipulation plane.

## 3. Results and discussions

In this section, a proof-of-principle experiment has been carried to demonstrate optical centrifuge. Firstly, we design its hologram, as shown in Fig. 2a, and its reconstructed phase profile and intensity distribution at manipulation plane are presented in Fig. 2b-c, respectively. As seen from Fig. 2b, its phase profile is highly desirable for optical centrifuge, which highly meets our design predict. For Fig. 2c, its reconstructed intensity is approximately uniform at large area, except that the intensity at small area near center point is relatively larger. In this case, it don't substantially affect the centrifuge performance in the experiment, because its intensity gradient force is weak. The measured intensity distribution is provided in Fig. 2d, which is almost identical to its reconstructed intensity in Fig. 2c. It reveals that our hologram design has high accuracy to generate its intensity and phase profile at manipulation plane.

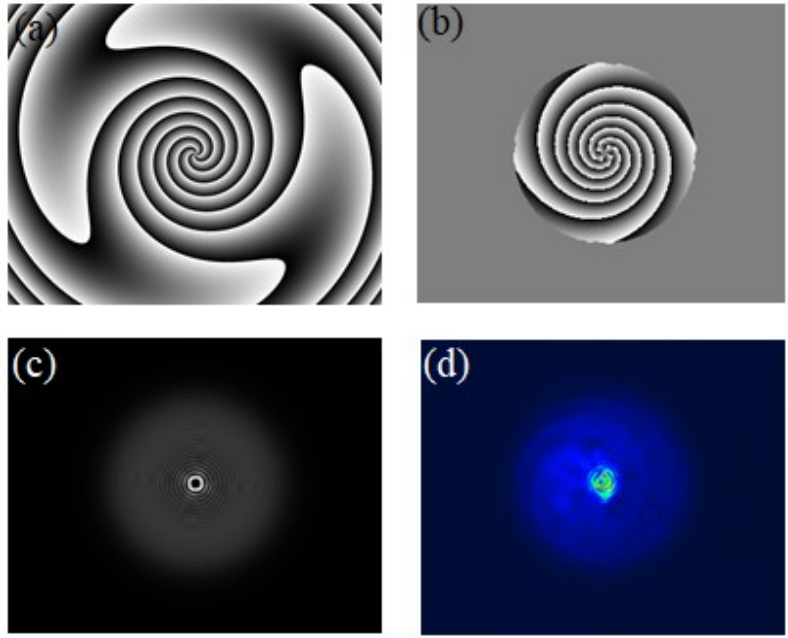

**Fig. 2.** The optical centrifuge by using phase gradient force. (a) The designed hologram. (b) The reconstructed phase profile. (c) The reconstructed intensity. (d) The measured intensity distribution.

In the following, Au nanoparticles with diameter of 200 nm are employed to demonstrate centrifuge motion in holographic optical tweezer system, which can be directly used for optical clearing. The position trajectory of nanoparticle is given in Figure 3(a), whose motion is driven by using holograms in Figure 2(a). It shows that it obviously moves in a centrifuge mode, which is strongly preferred as an optical clearing. Here it needs to note that the particle is assumed to start to move at the origin point of coordinate. Then, we provide its trajectories varying with time in polar coordinate system, as given in Figure 3(b)-(c), respectively. Similarly, its starting point is assumed to be origin point in polar coordinate system. Here, it should be mentioned that the dashed line in Figure 3(b)-(c) denotes the time point of starting to move in centrifuge motion. Importantly, it indicates that the particle rapidly moves away in a short time, as shown in Figure 3(b)-(c), which exhibits an excellent performance in centrifuge motion. Through calculation, its maximum velocity reaches over 120 μm/s and its average velocity is about 29 μm/s, which is highly desired while used as optical clearing.

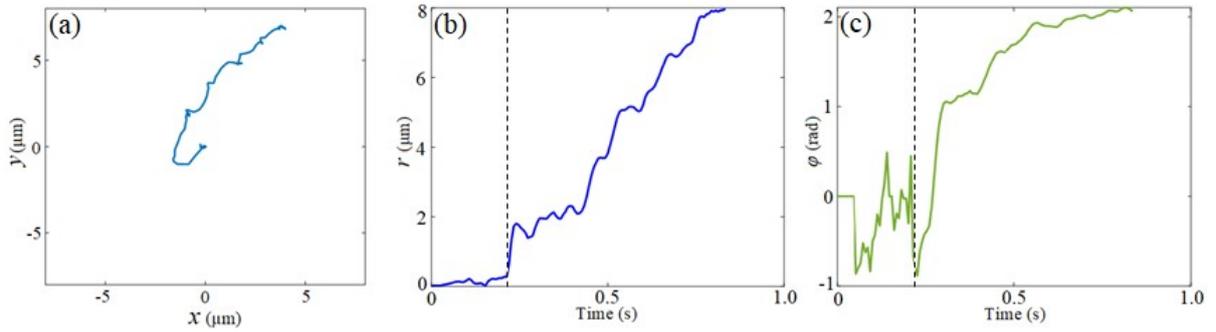

**Fig. 3.** The optical centrifuge motion. (a) The position trajectories of nanoparticle. (b) The radial position variation and (c) angle variation with time in polar coordinate system.

Optical separation is an important function in optical manipulation [28-29]. Herein, we show an optical separation of nanoparticles with different refractive index, by using optical centrifuge. Firstly, the mixed nanoparticles of Au and polystyrene nanosphere with diameter of 500 nm are trapped by using a focused Gaussian beam, as named stage I. Then, the mixed particles are driven in centrifuge mode by using hologram in Figure 2(a), as called stage II. After the Au nanoparticles are separated from the mixed particles, the third hologram is utilized for respectively positioning the trapped Au and polystyrene nanosphere at stage III. The position trajectories of three Au nanoparticles in optical sorting experiment is given in Figure 4(a), which indicates that three Au nanoparticles can be well driven to the desired position. Additionally, their radial position variation with time is shown in Figure 4(b), in which stage I, II and III at the different time are marked. It finds that all particles run away from center to edge in a short time (about 0.25s). In contrast, the polystyrene nanospheres almost stay the original position, due to fact that its driving force is small and the time is very short at stage II, whose position trajectories and radial position variation with time are not provided here. The snapshots of nanoparticle motion at different time are presented in Figure 4 (c), in which the polystyrene nanospheres are marked by red arrows. It demonstrates that our proposal for optical separation can well separate Au nanoparticles from mixed nanoparticles in a short time, which largely broaden the capability of conventional centrifuge.

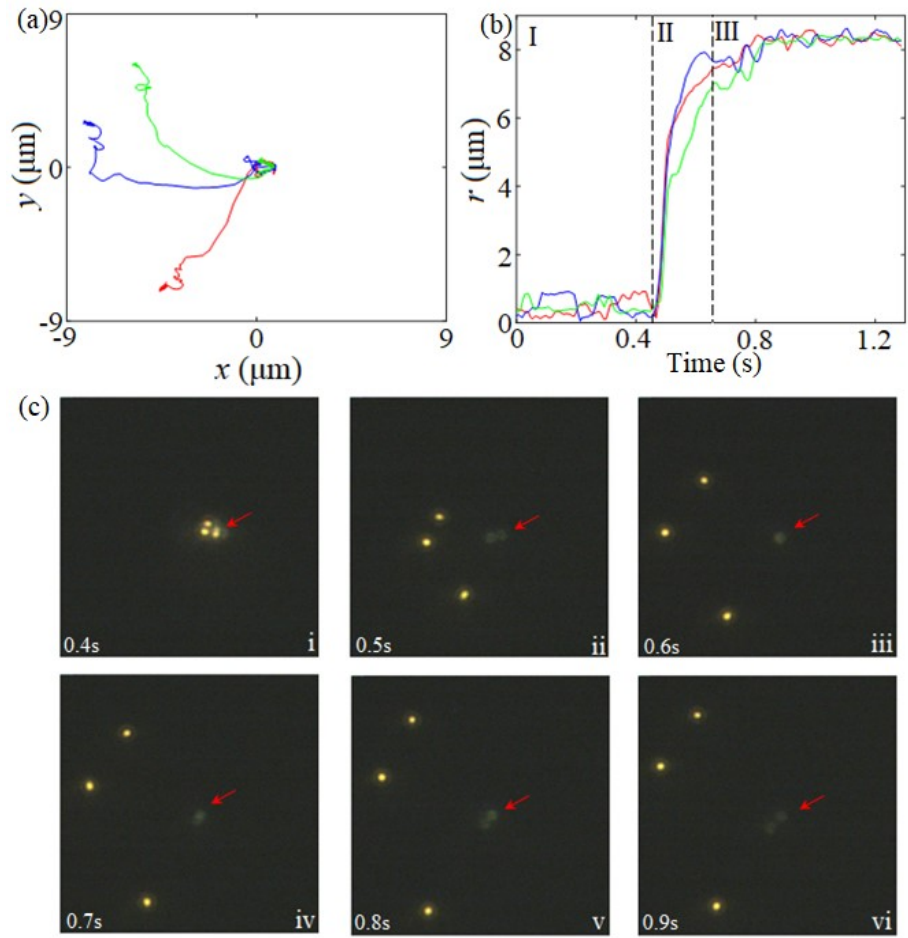

**Fig. 4.** The optical separation. (a) the position trajectories of three nanoparticles, (b) the radial position of three nanoparticles in different time, (c) the snapshots of nanoparticle motion at different time.

## 4. Conclusion

In this work, we have proposed and experimentally demonstrated optical centrifuge by employing optical lateral force arising from phase gradient profiles. The optical centrifuge has remarkable advantages such as rapid speed, flexible design and easy control. The reconstructed intensity and phase profiles are in consistent with our desirable design predict. Importantly, the proof-of principle experiments show that the optical centrifuge can easily achieve rapid operation of optical clearing and separation. For instance, its velocity in optical centrifuge motion can reach maximum in a short time, which is over 120 μm/s. More importantly, the proposed optical centrifuge provides a new tool for developing new manipulation function including optical clearing, optical separation and controllable collective transport.